\documentclass[aps,prl,twocolumn,groupedaddress,floatfix,showpacs]{revtex4}

\usepackage{graphicx}% Include figure files
\usepackage{dcolumn} % Align table columns on decimal point
\usepackage{bm}      % bold math
\usepackage{amsmath} % \text{}

\begin{document}

\title{Charge and orbital order in Fe$_3$O$_4$}

\author{I.~Leonov$^{1,2}$ \email[e-mail: ]{Ivan.Leonov@Physik.uni-Augsburg.de},
A.~N.~Yaresko$^3$, V.~N.~Antonov$^4$, M.~A.~Korotin$^2$, and V.~I.~Anisimov$^2$}
\address{$^1$ Theoretical Physics III, Institute for Physics, University of Augsburg, Germany}
\address{$^2$ Institute of Metal Physics, Russian Academy of Science-Ural Division, 
620219 Yekaterinburg GSP-170, Russia}
\address{$^3$ Max-Planck Institute for the Physics of Complex Systems, Dresden, Germany}
\address{$^4$ Institute of Metal Physics, Vernadskii Street, 03142 Kiev, Ukraine}

\date{\today}

\begin{abstract}
Charge and orbital ordering in the low-temperature monoclinic structure of magnetite
(Fe$_3$O$_4$) is investigated using LSDA+$U$. While the difference between $t_{2g}$ 
minority occupancies of Fe$^{2+}_B$ and Fe$^{3+}_B$ cations is large and gives 
direct evidence for charge ordering, the screening is so effective that the total 
$3d$ charge disproportion is rather small. The charge order has a pronounced [001] 
modulation, which is incompatible with the Anderson criterion. The orbital order 
agrees with the Kugel-Khomskii theory.
\end{abstract}


\maketitle

%%%%%%%%%%%%%%%%%%%%%%%%%%%%%%%%%%%%%%%%%%%%%%%%%%%%%%%%%%%%%%%%%%%%%%%%%%%%%%%%%
%{\large \bf Introduction}
%%%%%%%%%%%%%%%%%%%%%%%%%%%%%%%%%%%%%%%%%%%%%%%%%%%%%%%%%%%%%%%%%%%%%%%%%%%%%%%%%

The magnetic properties of magnetite, the famous lodestone, has fascinated 
mankind for several thousand years already \cite{Mattis}. Even today, in 
view of the possible technological importance of this material for 
spintronics \cite{CBBPP98}, and because of the still not well understood 
low-temperature properties, magnetite remains at the focus of active 
research.

Magnetite is a mixed-valent system and is the parent compound for
magnetic materials such as maghemite (Fe$_2$O$_3$) and spinel ferrites. 
At room temperature it crystallizes in the inverted cubic spinel structure 
$Fd\bar{3}m$ with tetrahedral $A$-sites occupied by Fe$^{3+}$ cations, 
whereas octahedral $B$-sites are occupied by an equal number of randomly 
distributed Fe$^{2+}$ and Fe$^{3+}$ cations. Magnetite is ferrimagnetic 
with an anomalous high critical temperature T$_C \sim 860$ K, the 
$A$-site magnetic moments being aligned antiparallel to the $B$-site 
moments. At room temperature Fe$_3$O$_4$ is a poor metal with an 
electronic conductivity of 4 m$\Omega$ cm. Upon further cooling 
a first-order metal-insulator (Verwey-) transition occurs at 
T$_V \sim 120$ K where conductivity abruptly decreases by two orders 
of magnitude. According to Verwey  this transition is caused by the 
ordering of Fe$^{2+}$ cations on the $B$ sublattice, with a simple 
charge arrangement of $(001)_c$ planes (indexed on the cubic cell) 
alternately occupied by 2+ and 3+ Fe$_B$ cations (Verwey charge 
ordering model) \cite{V39,VHR47}. This particular charge order (CO) 
obeys the so-called Anderson criterion \cite{And56} for minimal 
electrostatic repulsion leading to a short range CO pattern, namely 
tetrahedra of $B$-sites with an equal number of 2+ and 3+ cations. 
Since then a wide range of other CO models has been proposed which, 
however, all make use of the Anderson criterion \cite{Miz78,ZSP90}.

Later experiments showed that the Verwey transition is accompanied by 
a structural distortion from cubic to the monoclinic structure which 
has not been fully resolved so far \cite{IKS82,ZSP90}. Although the 
experiments rule out the Verwey CO below T$_V$ all attempts to 
construct a refined CO model from neutron-scattering data set failed 
because neutron scattering is more sensitive to atomic displacements 
than to charge ordering. In the absence of a definitive, experimentally 
determined structure many theoretical models for the low-temperature 
(LT) phase of magnetite \cite{Rev01} have been proposed. They include 
purely electronic \cite{CC71,CC73} and electron-phonon \cite{Mot80,Y80} 
models for CO, as well as a bond dimerized ground state without charge 
separation \cite{SOF+02}. In particular all previous LDA$(+U)$ 
calculations were performed for undistorted cubic unit cell. In spite 
of the fact that the amplitudes of these distortions are quite small 
this approximation for the LT unit cell in LSDA+$U$ calculations 
results inevitably in the Verwey CO. This problem is overcome in our 
work using recently refined crystal structure \cite{WAR01,WAR02} in 
which the ground state with more complicated CO was found. This also 
confirms Szotek $et~al.$ \cite{STSPSW} conclusion that the Verwey CO 
is not the ground state for magnetite, which is based on 
self-interaction corrected local spin density calculations for the 
refined LT crystal structure.

In this paper we report LSDA+$U$ calculations \cite{ldau:AZA91,ldau:LAZ95} in
the tight-binding linear muffin-tin orbital (TBLMTO) calculation scheme for
Fe$_3$O$_4$ in the $P2/c$ structure \cite{WAR01,WAR02}. Motivated by our 
results, we propose an order parameter, defined as the difference between 
$t_{2g}$ minority spin occupancies of Fe$^{2+}_B$ and Fe$^{3+}_B$ cations. 
This order parameter is found to be quite large, although the total $3d$ 
charge difference between these cations, is small. It seems certain that 
magnetite is long-range ordered below T$_V$, in contrast to the intermediate 
valence regime proposed by Garc{\'\i}a $et~al.$ \cite{GSP00,GSP01,GSBP}. 

%%%%%%%%%%%%%%%%%%%%%%%%%%%%%%%%%%%%%%%%%%%%%%%%%%%%%%%%%%%%%%%%%%%%%%%%%%%%%%%%%
%{\large \bf Crystal structure}
%%%%%%%%%%%%%%%%%%%%%%%%%%%%%%%%%%%%%%%%%%%%%%%%%%%%%%%%%%%%%%%%%%%%%%%%%%%%%%%%%

Recently, the LT superstructure of magnetite was refined by Wright $et~al.$ 
\cite{WAR01,WAR02}. The space group was confirmed to be monoclinic $Cc$, but 
the structure refinement was only possible in $P2/c$ group. They proposed the 
$a_c / \sqrt{2} \times a_c/ \sqrt{2} \times 2 a_c$ subcell ($a_c$ is the cell 
parameter of the undistorted cubic unit cell) with $P2/c$ space group symmetry. 
The additional $Pmca$ orthorhombic symmetry constrains were also applied.
The refined cell parameters were $a= 5.94437(1)$~\AA, $b= 5.92471(2)$~\AA, 
$c= 16.77512(4)$~\AA, and $\beta= 90.236(1)^{\circ}$. The structural evidence 
for the CO below the transition in the refined crystal structure, which is 
based on estimations of the mean $B$-site-to-oxygen distances or the bond 
valence sum (BVS) analyses, was also declared \cite{WAR01,WAR02}. However, 
this refined structure analysis has recently been found to be controversial 
\cite{GSBP}. The lack of atomic long-range order and, as a result, the 
intermediate valence regime below the Verwey transition were proposed.
Indeed, a difference between averaged Fe-O distances of compressed and
expanded FeO$_6$ octahedra corresponds to the maximum limit of charge
disproportion ($0.2\bar{e}$), which has the same order as the total 
sensitivity (including experimental errors) of BVS method. This 
contradiction as well as the ambiguity of proposed CO schemes (two 
different CO classes were proposed: class-I in the $P2/c$ unit cell 
and class-II in the full $Cc$ superstructure), is resolved in our 
electronic structure study.

%%%%%%%%%%%%%%%%%%%%%%%%%%%%%%%%%%%%%%%%%%%%%%%%%%%%%%%%%%%%%%%%%%%%%%%%%%%%%%%%%
%{\large \bf Results: LSDA}
%%%%%%%%%%%%%%%%%%%%%%%%%%%%%%%%%%%%%%%%%%%%%%%%%%%%%%%%%%%%%%%%%%%%%%%%%%%%%%%%%

We perform LSDA and LSDA+$U$ calculations for Fe$_3$O$_4$ in the $P2/c$ 
structure. For simplicity we neglect the small spin-orbit coupling (in 
previous calculations for cubic Fe$_3$O$_4$ spin-orbital splitting of the 
$3d$-band was found to be two orders of magnitude smaller than the crystal 
field splitting \cite{AHA+01}). The LSDA calculations give only a 
half-metallic ferrimagnetic solution without CO. Partially filled bands at 
the Fermi level originate from the minority spin $t_{2g}$ orbitals of 
Fe$_B$ cations. The tetrahedral Fe$_A$ cations do not participate in the 
formation of bands near the Fermi level, since their minority and majority 
spin $3d$ states are completely occupied and completely empty, respectively.
Thus, the LSDA results qualitatively agree with previous band-structure
calculations for the cubic phase \cite{STSPSW,AEH+96,ZS91,YH99}.
Apparently, only crystal structure distortion from cubic to monoclinic 
phase is not sufficient to explain metal-insulator transition and charge 
ordering in magnetite. The electron-electron correlations, mainly in the 
$3d$ shell of Fe cations, play significant role.

%%%%%%%%%%%%%%%%%%%%%%%%%%%%%%%%%%%%%%%%%%%%%%%%%%%%%%%%%%%%%%%%%%%%%%%%%%%%%%%%%
%{\large \bf Results: LSDA+U, U}
%%%%%%%%%%%%%%%%%%%%%%%%%%%%%%%%%%%%%%%%%%%%%%%%%%%%%%%%%%%%%%%%%%%%%%%%%%%%%%%%%

\begin{figure}[tbp!]
\centerline{\includegraphics[width=0.4\textwidth,clip]{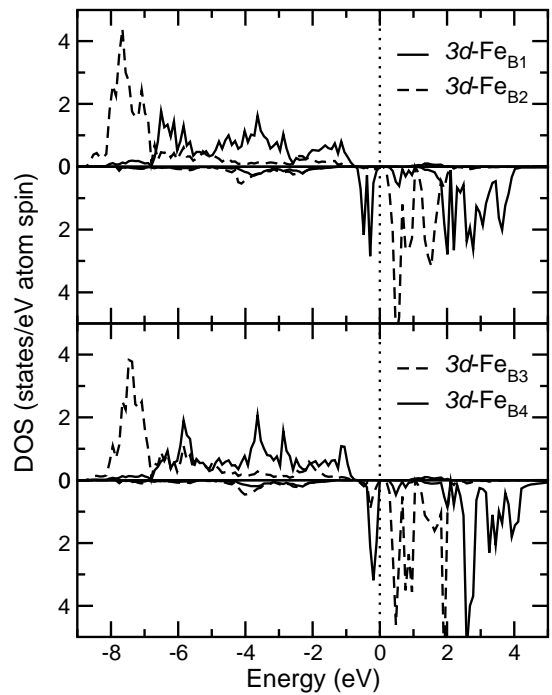}}
\caption{\label{fig:dos}
Partial DOS obtained from the LSDA+$U$ calculations with $U$=5 eV and $J$=1 eV 
for the low-temperature $P2/c$ phase of Fe$_3$O$_4$. The top of the valence band
is shown by dotted lines.}
\end{figure}

To proceed further we take into account the strong electronic correlations 
in Fe $3d$ shell using LSDA+$U$ method. In contrast to the LSDA, already 
with $U$ and $J$ obtained from constrained calculations (4.5 eV and 1 eV, 
respectively) a charge ordered insulator with energy gap of 0.03 eV was
obtained. On the other hand, the calculation of $U$ depends on theoretical 
approximations and, as a rule, the accuracy does not exceed 10-20\%. A 
reasonably good agreement of the calculated gap of 0.18 eV with the
experimental value of 0.14 eV \cite{PIT98} was found using the $U$ value 
of 5 eV. As shown in Fig.~\ref{fig:dos}, the gap opens between occupied 
and unoccupied $t_{2g \downarrow}$ states of Fe$_{B(1)}^{2+}$,
Fe$_{B(4)}^{2+}$ and Fe$_{B(2)}^{3+}$, Fe$_{B(3)}^{3+}$ respectively.
The remaining unoccupied Fe$_B$ states are pushed by strong Coulomb
repulsion to the energies above 2 eV.

The obtained CO is described by a dominant $[001]_c$ charge density
wave with a minor $[00\frac{1}{2}]_c$ modulation in the phase of CO
and coincides with the class-I CO, which was proposed earlier by
Wright $et~al.$ \cite{WAR01,WAR02}. Thus, the LSDA+$U$ calculations
confirm violation of the Anderson criterion for Fe$_3$O$_4$ in the LT
phase. In order to check the stability of the obtained CO solution we 
performed additional self-consistent LSDA+$U$ calculations both for the 
$P2/c$ structure and $Cc$ supercell. In the first case the Verwey CO was 
used as the starting CO model, while in the second one we started from 
class-II CO, shown in Fig.~2 of Ref.~\onlinecite{WAR02}. However, it was 
found that these CO models are unstable and the obtained self-consistent 
solutions coincide with the stable one found previously, i.e. for certain 
value of $U$ the obtained CO does not depend on initial charge arrangement. 
Obviously, we did not check all possible CO scenarios within $P2/c$ unit 
cell or $Cc$ supercell, but our results consistently indicate that the 
obtained class-I CO solution is the ground state of Fe$_3$O$_4$ in the LT 
phase. It is important to note that LSDA+$U$ calculations performed for an 
undistorted $P2/c$ supercell of $Fd\bar{3}m$ structure result in an 
insulating CO solution which \textit{is compatible} with the Verwey CO model.

%%%%%%%%%%%%%%%%%%%%%%%%%%%%%%%%%%%%%%%%%%%%%%%%%%%%%%%%%%%%%%%%%%%%%%%%%%%%%%%%%
%%%%%%%%%%%%%%%%%%%%%%%%%%%%%%%%%%%%%%%%%%%%%%%%%%%%%%%%%%%%%%%%%%%%%%%%%%%%%%%%%

%%%%%%%%%%%%%%%%%%%%%%%%%%%%%%%%%%%%%%%%%%%%%%%%%%%%%%%%%%%%%%%%%%%%%%%%%%%%%%%%%
\begin{table}[tbp!]
\caption{\label{tab:occ}Total and $l$-projected charges, magnetic moments, and
the occupation of the most populated $t_{2g}$ minority orbital calculated for
inequivalent Fe$_{B}$ ions in the low-temperature $P2/c$ phase of Fe$_3$O$_4$
\cite{orbs01}.}
\begin{ruledtabular}
\begin{tabular}{lccccccc}
Fe$_{B}$ ion & $q$ &$q_s$ &$q_p$ &$q_d$ & $M$ ($\mu_{\text{B}}$)
& $t_{2g \downarrow}$ orbital & $n$ \\
\hline
Fe$_{B(1)}$  & 6.04 & 0.17 & 0.19 & 5.69 & 3.50 &$d_{xz} \mp d_{yz}$& 0.76 \\
Fe$_{B(2)}$  & 5.73 & 0.19 & 0.21 & 5.44 & 3.94 &                   & 0.27 \\
Fe$_{B(3)}$  & 5.91 & 0.19 & 0.21 & 5.51 & 3.81 &                   & 0.24 \\
Fe$_{B(4)}$  & 6.03 & 0.16 & 0.18 & 5.69 & 3.48 & $d_{x^2-y^2}$     & 0.80 \\
\end{tabular}
\end{ruledtabular}
\end{table}
%%%%%%%%%%%%%%%%%%%%%%%%%%%%%%%%%%%%%%%%%%%%%%%%%%%%%%%%%%%%%%%%%%%%%%%%%%%%%%%%%

%%%%%%%%%%%%%%%%%%%%%%%%%%%%%%%%%%%%%%%%%%%%%%%%%%%%%%%%%%%%%%%%%%%%%%%%%%%%%%%%%
%{\large \bf Results: LSDA+U occupation matrix analysis}
%%%%%%%%%%%%%%%%%%%%%%%%%%%%%%%%%%%%%%%%%%%%%%%%%%%%%%%%%%%%%%%%%%%%%%%%%%%%%%%%%

An analysis of occupation matrices of $3d$-Fe$_B$ minority spin
states confirms substantial charge disproportionation. As shown
in Table~\ref{tab:occ}, one of the $t_{2g \downarrow}$ states of
Fe$^{2+}_{B(1)}$ and Fe$^{2+}_{B(4)}$ cations is almost completely
filled with the occupation numbers $n\approx 0.8$. On the other
hand, the remained two $t_{2g \downarrow}$ orbitals of the
Fe$^{2+}$ cations have significantly smaller population of about 0.15. 
The
occupation numbers of $t_{2g \downarrow}$ orbitals for
Fe$^{3+}_{B(2)}$ and Fe$^{3+}_{B(3)}$ do not exceed 0.3, which
gives the value of about 0.5 for the largest difference between 
Fe$^{2+}_{B}$ and Fe$^{3+}_{B}$ $t_{2g}$ minority spin populations. 
The corresponding total $3d$ charges difference
(0.23) and disproportionation of the total electron charges inside
the atomic spheres of Fe$^{2+}_{B}$ and Fe$^{3+}_{B}$ cations
(0.24) is in a reasonably good agreement with the value of 0.2
estimated from BVS analysis of monoclinic structure. This shows
that the change of the $t_{2g \downarrow}$ occupations caused by
the charge ordering is very effectively screened by the
rearrangement of the other Fe electrons.

Significant contribution
to the charge screening is provided by Fe$_B$ $e_g$ states.
Although the bands originating from these states are located well
above the energy gap, the $e_g$ minority orbitals form relatively
strong $\sigma$-bonds with 2$p$-states of the oxygen octahedron
and, as a result, give appreciable contribution to the occupied
part of the valence band. The above mentioned screening 
reduces the
energy loss due to the development of CO incompatible with the
Anderson criterion in the LT phase of Fe$_3$O$_4$. Hence, due to
strong screening effects, the order parameter,
introduced as the total $3d$ charge difference between 2+ and 3+ 
Fe$_B$
cations, is ill-defined. That explains why the BVS analysis
does not give convincing proof of CO existence. Apparently,
the well-defined order parameter is the difference of
$t_{2g \downarrow}$ occupancies for Fe$^{3+}_B$ and
Fe$^{2+}_B$ ions, which amounts to 50\% of ideal ionic CO model
and clearly pronounces the existence of CO below the Verwey
transition.

%%%%%%%%%%%%%%%%%%%%%%%%%%%%%%%%%%%%%%%%%%%%%%%%%%%%%%%%%%%%%%%%%%%%%%%%%%%%%%%%%
%{\large \bf Results: LSDA+U orbital ordering}
%%%%%%%%%%%%%%%%%%%%%%%%%%%%%%%%%%%%%%%%%%%%%%%%%%%%%%%%%%%%%%%%%%%%%%%%%%%%%%%%%

%%%%%%%%%%%%%%%%%%%%%%%%%%%%%%%%%%%%%%%%%%%%%%%%%%%%%%%%%%%%%%%%%%%%%%%%%%%%%%%%%
\begin{figure}[tbp!]
\centerline{\includegraphics[width=0.325\textwidth,clip]{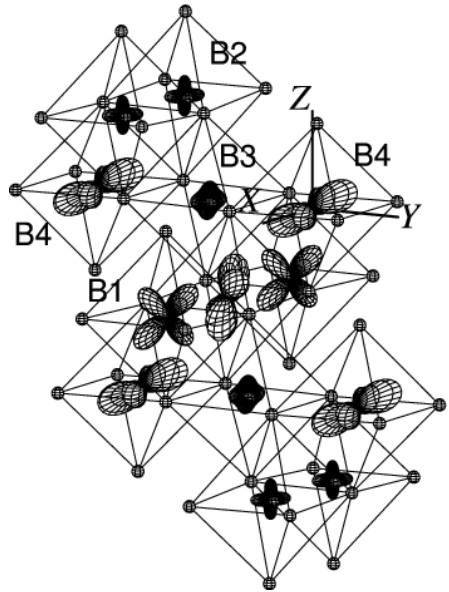}}
\caption{\label{fig:orb} The LSDA+$U$ angular distribution of the minority 
spin $3d$ electron density of Fe$_B$ cations for the low-temperature $P2/c$ 
phase of Fe$_3$O$_4$. The angular distribution is calculated according to
$\rho(\theta,\phi)=\sum_{m,m'}n_{m,m'}~Y^\ast_{m}(\theta,\phi)~Y_{m'}(\theta,\phi)$,
where $n_{m,m'}$ is the occupation matrix of $d$ minority states for Fe$_B$ 
atoms. $Y_{m}(\theta,\phi)$ are corresponding spherical harmonics. Oxygen atoms 
are shown by small spheres. The size of orbital corresponds to its occupancy.}
\end{figure}
%%%%%%%%%%%%%%%%%%%%%%%%%%%%%%%%%%%%%%%%%%%%%%%%%%%%%%%%%%%%%%%%%%%%%%%%%%%%%%%%%

The self-consistent solution obtained from the LSDA+U calculations
is not only charge but also orbitally ordered. The occupied minority 
spin $t_{2g}$ state of Fe$^{2+}_{B(1)}$ and Fe$^{2+}_{B(4)}$ cations 
is predominantly of $d_{xz}\pm d_{yz}$ and $d_{x^2-y^2}$ character, 
respectively. This is illustrated by Fig.~\ref{fig:orb} which shows 
the angular distribution of the minority spin $3d$ Fe$_B$ electron 
density. Note, however, that the $P2/c$ frame is rotated by an angle 
$\sim \pi/4$ with respect to the cubic one and the angular dependence 
of $t_{2g}$ states is given by $d_{xy} \pm d_{yz}$ and $d_{x^2-y^2}$ 
combination of cubic harmonics. The obtained relative orientation of 
occupied Fe$_B$ $t_{2g}$ minority orbitals corresponds to the 
anti-ferro-orbital order. Since all Fe$_B$ cations are ferromagnetically 
coupled, the obtained orbital order is consistent with the anti-ferro-orbital 
ferromagnetic state, which is the ground state of the degenerate Hubbard 
model according to the Kugel-Khomskii theory \cite{KH75}. This orbital 
order leads to the corresponding distortions of FeO$_6$ octahedra.
An analysis of interatomic distances in the monoclinic structure
(Table~\ref{tab:dist}) shows that the average Fe$_{B(1)}$--O distance (2.087
\AA) in the plane perpendicular to one of the diagonals of the distorted
Fe$_{B(1)}$O$_6$ octahedron is considerably larger than average distances in
the other two planes (2.063 and 2.067 \AA). It turns out that the occupied
Fe$_{B(1)}$ $t_{2g}$ minority spin orbital is the one oriented in the plane
with the largest average Fe$_{B(1)}$--O distance. The same is also true for
Fe$_{B(4)}$ ion but in this case the variation of the average Fe$_{B(4)}$--O
distances is smaller (2.074 \AA\ vs.\ 2 $\times$ 2.067 \AA) and, as a
consequence, the out-of-plane rotation of the occupied $t_{2g \downarrow}$
orbital is stronger.

%%%%%%%%%%%%%%%%%%%%%%%%%%%%%%%%%%%%%%%%%%%%%%%%%%%%%%%%%%%%%%%%%%%%%%%%%%%%%%%%%
\begin{table}[tbp!]
\caption{\label{tab:dist}The averaged Fe$_B$--O distances in the plane of 
$t_{2g}$ orbitals for $P2/c$ structure of Fe$_3$O$_4$ \cite{orbs01}}
\begin{ruledtabular}
\begin{tabular}{lcccc}
Fe$_B$ atom & orbital & $d_{\mathsf{orb.}}$ (\AA) & $d_{\mathsf{av.}}$ (\AA)\\
\hline
Fe$_{\mathsf{B(1a)}}$ & $d_{xz}+d_{yz}$ & 2.067 & 2.072  \\
                      & $d_{xz}-d_{yz}$ & 2.087 &        \\
                      & $d_{x^2-y^2}$   & 2.063 &        \\
Fe$_{\mathsf{B(1b)}}$ & $d_{xz}+d_{yz}$ & 2.087 & 2.072  \\
                      & $d_{xz}-d_{yz}$ & 2.067 &        \\
                      & $d_{x^2-y^2}$   & 2.063 &        \\
Fe$_{\mathsf{B(4)}}$  & $d_{xz} \pm d_{yz}$ & 2.067 & 2.069  \\
                      & $d_{x^2-y^2}$   & 2.074 &        \\
\end{tabular}
\end{ruledtabular}
\end{table}
%%%%%%%%%%%%%%%%%%%%%%%%%%%%%%%%%%%%%%%%%%%%%%%%%%%%%%%%%%%%%%%%%%%%%%%%%%%%%%%%%

%%%%%%%%%%%%%%%%%%%%%%%%%%%%%%%%%%%%%%%%%%%%%%%%%%%%%%%%%%%%%%%%%%%%%%%%%%%%%%%%%
%{\large \bf Discussion}
%%%%%%%%%%%%%%%%%%%%%%%%%%%%%%%%%%%%%%%%%%%%%%%%%%%%%%%%%%%%%%%%%%%%%%%%%%%%%%%%%

As was shown earlier, the Verwey CO model possesses the minimum electrostatic
repulsion energy among all possible CO models \cite{ZSP90}. But due to the 
existence of two perpendicular families of $B$-site chains (for instance 
$[110]_c$ and $[1\bar{1}0]_c$) correspondingly occupied by 2+ and 3+ Fe$_B$ 
cations the lattice ``feels'' significant stresses and tends to expand in 
one ($[110]_c$) and to compress in the another ($[1\bar{1}0]_c$) direction.
Therefore, the Verwey CO gives a significant ``elastic'' energy contribution 
into the total energy and in spite of the lowest electrostatic energy becomes 
less favorable than other arrangements. The competition of these two (elastic 
and electrostatic) contributions in the total energy appears to be responsible 
for the charge order, which is realized in the experimentally observed 
low-temperature monoclinic structure. In this CO scheme the alternating 
$(001)_c$ planes occupied by 2+ (``occupied'' plane) and by 3+ (``empty'' 
plane) Fe$_B$ cations are separated by the ``partially'' occupied plane. 
This $(001)_c$ planes order makes the difference between $[110]_c$ and 
$[1\bar{1}0]_c$ directions less pronounced and significantly reduces the 
lattice stress and, as a result, reduces the elastic energy contribution in 
the total energy. We propose that this scenario is the primary cause for 
development of the class-I CO found below the Verwey transition.

%%%%%%%%%%%%%%%%%%%%%%%%%%%%%%%%%%%%%%%%%%%%%%%%%%%%%%%%%%%%%%%%%%%%%%%%%%%%%%%%%
%{\large \bf Summary}
%%%%%%%%%%%%%%%%%%%%%%%%%%%%%%%%%%%%%%%%%%%%%%%%%%%%%%%%%%%%%%%%%%%%%%%%%%%%%%%%%

In summary, in the present LSDA+$U$ study of the LT $P2/c$ 
phase of Fe$_3$O$_4$ we found a charge and orbitally ordered insulator with
an energy gap of 0.18 eV. This is in a good agreement with the experimental
value of 0.14 eV \cite{PIT98}. The obtained charge ordered ground state
is described by a dominant $[001]_c$ charge density wave with a minor
$[00\frac{1}{2}]_c$ modulation. The CO coincides with the earlier proposed 
class-I CO \cite{WAR01,WAR02} and confirms a violation of the Anderson
criterion \cite{And56}.
While the screening of the charge disproportion is so effective that the total
$3d$ charge disproportion is rather small (0.23), the charge order is well
pronounced with an order parameter defined as a difference of $t_{2g
\downarrow}$ occupancies of 2+ and 3+ Fe$_B$ cations (0.5).
This agrees well with the result of BVS analysis for monoclinic structure 
(0.2). The orbital order is in agreement with the Kugel-Khomskii theory 
\cite{KH75} and corresponds to the local distortions of oxygen octahedra 
surrounding Fe$_{B}$-sites.

We are grateful to D.~Schrupp, R.~Claessen, J.~P.~Attfield, P.~Fulde, 
and D.~Vollhardt for helpful discussions. The present work was supported 
by RFFI Grant No. 01-02-17063 and by DFG through Grant No. 484.

%\bibliography{jprb}

\end{document}